\documentclass{ragtime} 

\title[Modified Newtonian potentials for particles and fluids]%
      {Modified Newtonian potentials for particles and fluids in permanent 
rotation around black holes}

\author[V. Karas 
        and M. A. Abramowicz]   
       {Vladim\'{\i}r Karas\at[]{1,a} 
        and Marek A. Abramowicz\at[]{2,3,4}\\
        \ins{1}Astronomical Institute,
        Academy of Sciences of the Czech Republic,\splitins[1]
        Bo\v{c}n\'{\i} II 1401, CZ-141\,00 Prague,
        Czech Republic\\
        \ins{2}Copernicus Astronomical Center, Polish Academy of Sciences,\splitins[1]
        ul. Bartycka 18, P-00\,716 Warsaw, Poland\\
        \ins{3}Institute of Physics, Faculty of Philosophy and Science, Silesian University in Opava,\splitins[3] 
        Bezru\v{c}ovo n\'am.~13, CZ-746\,01 Opava, Czech Republic\\%
        \ins{4}Physics Department, Gothenburg University, SE-412\,96 G\"{o}teborg, Sweden\\%
        \ins{a}\Email{E-mail:vladimir.karas@cuni.cz}} 

\begin{document}
\newcommand{\dg}{{^{\circ}}}
\newcommand{\rg}{{r_{\rm{g}}}}
\newcommand{\rh}{{r_+}}
\newcommand{\rmb}{{r_{\rm{mb}}}}
\newcommand{\rms}{{r_{\rm{ms}}}}
\newcommand{\Phipw}{{\Phi_{_{\rm{PW}}}}}
\newcommand{\Phiabn}{{\Phi_{_{\rm{ABN}}}}}
\newcommand{\Phinw}{{\Phi_{_{\rm{NW}}}}}
\newcommand{\beq}{\begin{equation}}
\newcommand{\eeq}{\end{equation}}
\def\spose#1{\hbox to 0pt{#1\hss}}
\def\lta{\mathrel{\spose{\lower 3pt\hbox{$\mathchar"218$}}
     \raise 2.0pt\hbox{$\mathchar"13C$}}}
\def\gta{\mathrel{\spose{\lower 3pt\hbox{$\mathchar"218$}}
     \raise 2.0pt\hbox{$\mathchar"13E$}}}

\begin{abstract}
Modified Newtonian potentials have been proposed for the description of 
relativistic effects acting on particles and fluids in permanent orbital motion 
around black holes. Here we further discuss spherically symmetric potentials 
like the one proposed by Artemova, Bj\"ornson \& Novikov (1996, Astrophysical 
Journal, 461, 565), and we illustrate their virtues by studying the 
acceleration along circular trajectories. We compare the results with exact 
expressions in the spacetime of a rotating (Kerr) black hole.
\end{abstract}

\section{Introduction}\label{intro}
The motion of material around black holes, both particles and fluids, is of 
particular importance for the present-day models of some astronomical objects, 
such as galactic X-ray sources and active galactic nuclei. In these systems, 
matter may be found rather close to the black-hole horizon, at a few 
gravitational radii ($\rg=2GM/c^2;$ where $M$ is the mass of the central black 
hole), and the effects of general relativity on the motion must be taken into 
account. (We will here set $c=G=1$; in addition, we will measure lengths in 
units of $M$, so that $\protect\rg=2$ hereafter.) The relevant framework for 
discussing such fluids is then the Kerr spacetime of a rotating black hole (we 
consider here only test particles and fluids around the black hole, as is often 
done for astrophysical situations; however, for exact solutions of the Einstein 
equations with rotating bodies, see e.g.\ Islam 1985, and for more 
astrophysically realistic numerical solutions with self-gravitating material, 
see Lanza 1992; Nishida \& Eriguchi 1994). The relativistic effects for this 
matter can be ascribed mainly to two characteristic properties of motion around 
black holes: (i)~presence of the marginally stable orbit ($r=\rms$) and the 
marginally bound orbit ($r=\rmb$) which determine the regions of stable and 
energetically bound motion (their location determines also the inner edge of 
the toroidal fluid configurations; the exact location of $\rmb$ and $\rms$ can 
be found by studying the effective potential; see Bardeen, Press \& Teukolsky 
1972); (ii)~the frame dragging of non-equatorial orbits (Lense-Thirring 
precession, often called the Wilkins [1972] effect in the case of motion close 
to a Kerr black hole). In order to incorporate these effects within the 
Newtonian framework (which is of course technically easier than a fully 
relativistic self-consistent approach), numerous authors have adopted the 
original idea of Paczy\'nski \& Wiita (1980) and employed modifications of the 
Newtonian potential (Nowak \& Wagoner 1991; Artemova, Bj\"ornson \& Novikov 
1996; Crispino et al.\ 2011).\footnote{A different approach was adopted in 
Keres (1967); Israel (1970); and de Felice (1980), where some properties of the 
Kerr metric are described in terms of an axially symmetric (non-spherical) 
potential which reflects the asymptotic properties of test particle motion.}

In this Note we want to discuss simple (spherically symmetric) potentials 
appropriate for the description of matter in purely rotational motion, 
neglecting frame-dragging effects. This simplifies our discussion; see 
Semer\'ak \& Karas (1999) for detailed discussion and references concerning how 
to modify the Newtonian potential for including the effects of dragging. In 
previous studies, the main concern was about how to reproduce correctly the 
marginally bound and marginally stable orbits, since the properties of fluid 
tori are sensitive to the location of {\em{both\/}} of these orbits (cf.\ 
Muchotrzeb \& Paczy\'nski 1981; Abramowicz et al.\ 1988; Kato, Honma \& 
Matsumoto 1988; Chakrabarti 1990). See also Tejeda \& Rosswog (2013), and 
Barausse \& Lehner (2013) for a recent discussion and new developments.

The Paczy\'nski-Wiita potential, $\Phipw=-1/(r-\rg)$, reproduces the correct 
location of $\rmb$ and $\rms$ for a non-rotating black hole. Another form of 
the modified potential around a non-rotating black hole was used by Nowak \& 
Wagoner (1991) to study relativistic wave-modes in accretion discs: 
$\Phinw=-r^{-1}+3r^{-2}-12r^{-3}$ reproduces $\rms$ and the epicyclic frequency 
of radial oscillations $\kappa$. These two potentials $\Phipw$ and $\Phinw$ are 
not however applicable in the case of a rotating black hole. This situation has 
been treated by several authors, most recently and successfully by Artemova et 
al.\ (1996). Here we will further discuss the form of the potential which 
appears most convenient for modelling tori around rotating black holes. Note 
that tori rotate with non-Keplerian orbital velocity and they may extend well 
out of equatorial plane (Frank, King \& Raine 1992). One thus needs to consider 
also accelerated motion, though still in permanent rotation about the common 
axis of the black hole.

\section{Modified Newtonian potential for rotating black holes}
\subsection{Motivation}
The need for a practical and accurate modified potential leads to
constraining its form according to the following conditions:

\begin{description}
\item{(i)}~The modified potential should be a {\em{simple\/}} scalar function 
 of the spherical radius $r$;
\item{(ii)}~The potential should reduce to $\Phipw$ in the limit of zero
 rotation (black-hole angular momentum parameter $a=0$);
\item{(iii)}~The locations of $\rms$ and $\rmb$ should be correctly reproduced
 {\em{both\/}} for the non-rotating case ($a=0$, $\rms=6$, $\rmb=4$) and for
 the extreme rotating case ($a=1$, $\rmb=\rms=1$).
\end{description}
These requirements are satisfied by the function
\beq
\Phi=-\frac{1}{\left(r-\rh\right)^{\bar{\beta}}r^{1-\bar{\beta}}},
\label{our}
\eeq
where $\rh=1+\sqrt{1-a^2}$ is the black hole outer horizon, and the parameter 
$\bar{\beta}(a)$ is a free function. The choice for this is constrained by 
imposing that the values of $\rmb$ and $\rms$ should be exact in the 
Schwarzschild case and in the extreme Kerr case. We then adopt the simplest 
linear form: $\bar{\beta}=1-a$. Although, for $\bar{\beta}=0$, eq.\ (\ref{our}) 
reduces to the Newtonian potential which does not have marginally bound and 
marginally stable orbits, the correct location of $\rmb=\rms=1$ is nevertheless 
obtained in the limit of $a\rightarrow1$ (the extreme Kerr case).

One can verify that the properties of the potential (\ref{our}) are almost 
identical with those of the potential $\Phiabn$ of Artemova et al.\ (1996) 
which corresponds to their eq.\ (13) for the force:
\beq
F_5=-\frac{1}{r^{2-\beta}\left(r-\rh\right)^\beta},
\label{f5}
\eeq
with $\beta=(\rms/\rh)-1$. Expression (\ref{f5}) follows from the following 
conditions:

\begin{description}
\item{(i)}~The free-fall acceleration has a similar form to that for a 
Schwarzschild black hole;
\item{(ii)}~The free-fall acceleration diverges to infinity near $r=\rh$.
\item{(iii)}~The marginally stable orbit is reproduced {\em{exactly for
all\/}} values of $a$ ($0\leq{a}\leq1$).
\end{description}
Although the position of the important orbit $r=\rmb$ is not mentioned in the 
derivation of $F_5$, one can verify that the correct sequence is maintained for 
all $a$: $\rh\leq\rmb\leq\rms\leq6$ (indeed, the accuracy is very good as we 
will see in the next paragraph). The potential corresponding to $F_5$ is
\beq
\Phiabn=\frac{1}{(1-\beta)\rh}\left(1-\frac{\rh}{r}\right)^{1-\beta}
-\Phi_\infty,
\label{no5}
\eeq
with $\Phi_\infty=(1-\beta)^{-1}\rh^{-1}$.

\begin{figure}[tbh]
 \centering
 \includegraphics[width=\textwidth]{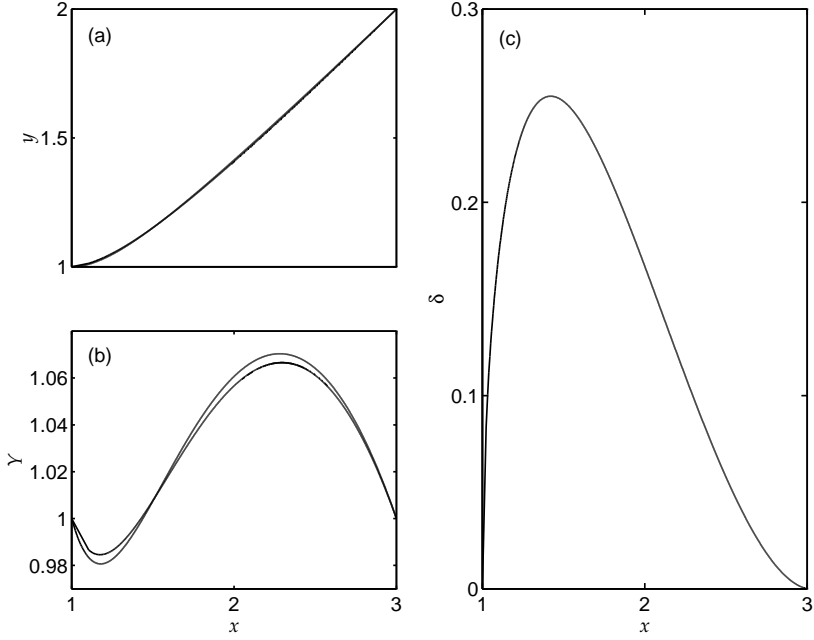}
 \caption{Radii of important orbits in the modified Newtonian potential
 $\Phi$ compared with the Kerr case: 
 (a)~the radius of the marginally bound orbit $y$ plotted as a function of
 the radius of the marginally stable orbit $x$, both measured in units of the 
 black hole outer horizon radius $\protect\rh$; (b)~the normalized marginally 
 bound radius $Y(x)=fy$; (c)~the relative difference $\delta(x)$ between the
 modified Newtonian and Kerr cases (see text for definitions).
 \label{fig1}}
\end{figure}

\subsection{Acceleration along circular orbits}

\begin{figure}[tbh]
\centering
\includegraphics[width=1.05\textwidth]{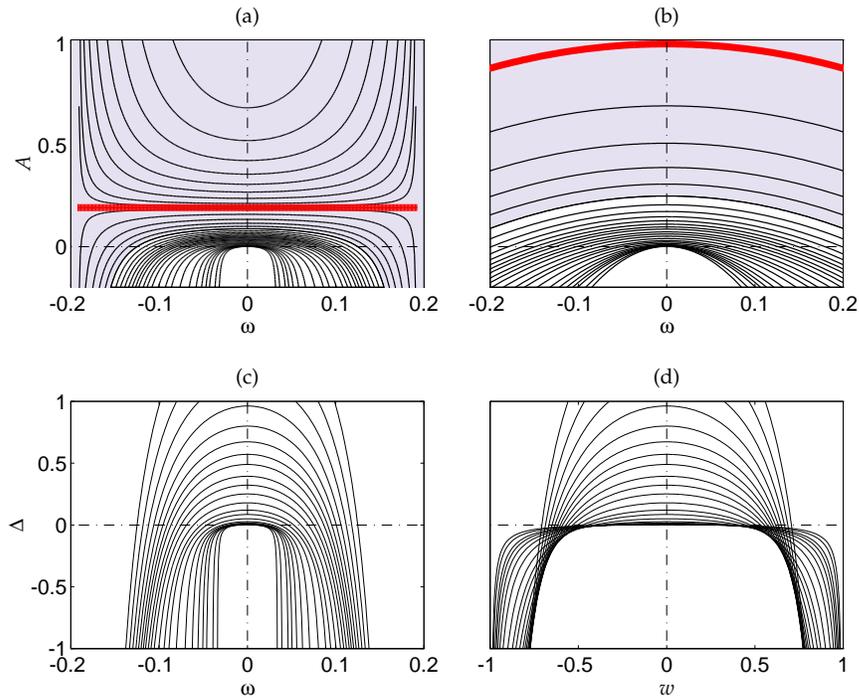}
 \caption{The acceleration magnitude $A$ is plotted as a function of the 
 angular velocity $\omega$ along circular orbits with different radii: (a)~the
 Schwarzschild case; (b)~the modified Newtonian case $\protect\Phipw$ 
 (the two plots are clearly different in the shaded
 area corresponding to $r<\rmb$ but they are quite similar outside
 that region, i.e.\ in the bottom part of the plots); 
 (c) and (d) show the relative difference $\Delta$ between the two 
 cases for radii in the range $\rmb\leq{r}\leq15\rh$. 
 Acceptable accuracy of $|\Delta|\lta0.1$ corresponds 
 to $|w|\lta0.5$ and $r\gta1.9\rmb$.
 \label{fig2}}
\end{figure}

We will now argue that the results for purely rotational motion of fluids
in potentials (\ref{our}) and (\ref{no5}) should be extremely similar and
close to the exact relativistic treatment in the Kerr metric. This
conjecture can be illustrated in two steps: first we will see that
$\rms$ and $\rmb$ are well reproduced (for both $\Phi$ and $\Phiabn$),
and then we will study acceleration along non-Keplerian circular orbits
(relevant for modelling tori).

\begin{figure}[tbh]
 \centering
 \includegraphics[width=1.05\textwidth]{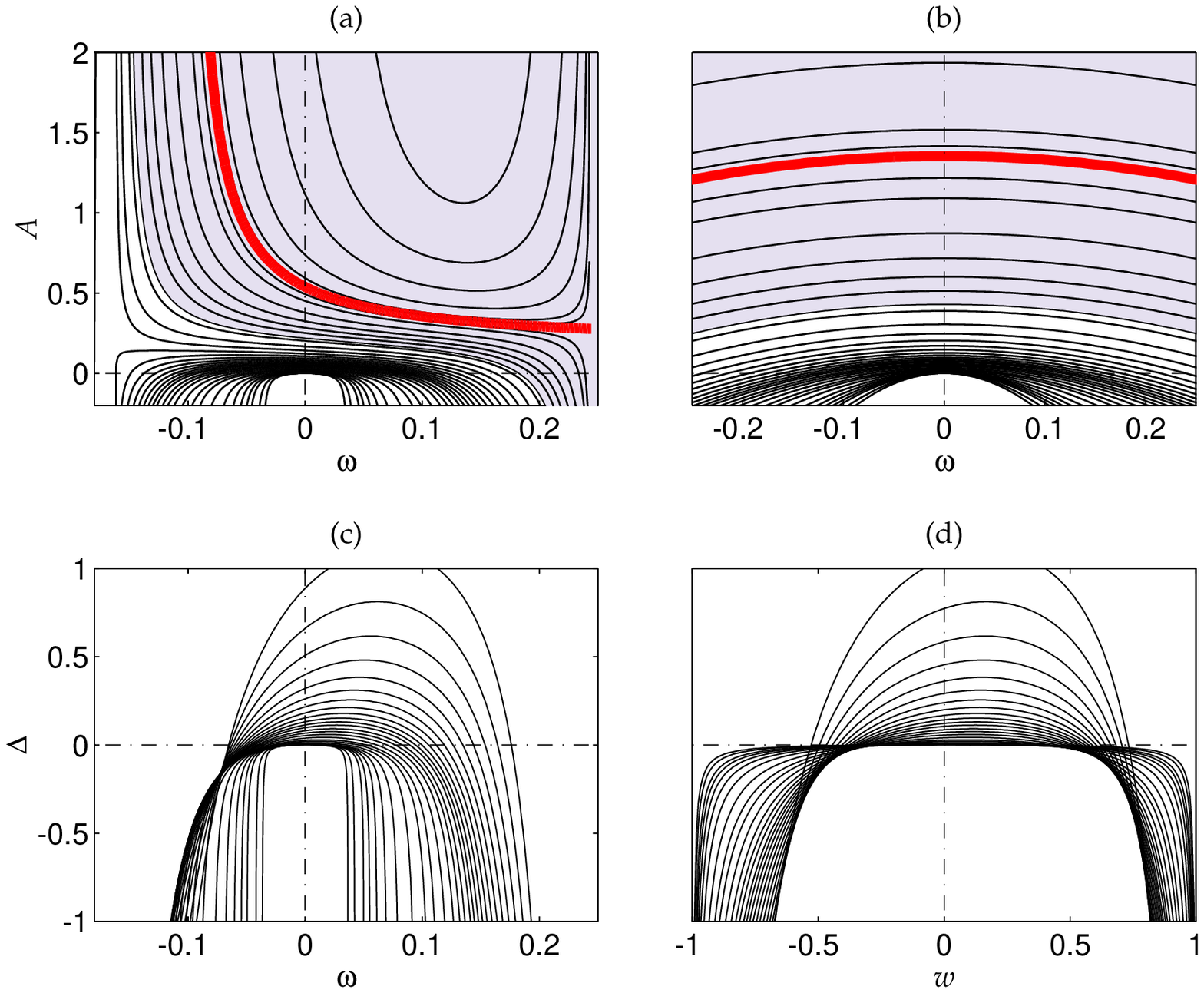}
  \caption{As in Fig.~\protect\ref{fig2} but for the Kerr $a=0.5$ case (a), and 
 for the equivalent $\protect\Phiabn$ case (b). Quite naturally, $|\Delta|$ is 
 on average large for the orbits with small radius. Comparing with analogous 
 graphs for $\protect\Phipw$, potential $\protect\Phiabn$ diminishes $|\Delta|$ 
 to smaller values, and is thus more accurate when $a$ is nonzero. The graph 
 here is not symmetrical about $\omega=0$ (due to frame-dragging in the Kerr 
 metric); accuracy is maintained to higher $|w|$ for corotating motion. Here, 
 an acceptable accuracy of $|\Delta|\lta0.1$ corresponds to $|w|\lta0.5$ and 
 $r\gta1.6\rmb$. \label{fig3}}
\end{figure}

We now illustrate the differences in the marginally bound radius as a function 
of the marginally stable radius. Fig.~\ref{fig1} compares our modified 
Newtonian ratios $y=\rmb/\rh$ and $x=\rms/\rh$ (evaluated by using eq.\ 
[\ref{our}]) with the corresponding values of the Boyer-Lindquist radial 
coordinate in the Kerr metric. In both cases $1\leq{x(a)}\leq3$ and 
$1\leq{y(a)}\leq2$ when the angular-momentum parameter varies in the range 
$1\geq{a}\geq0$. Fig.\ \ref{fig1}a shows that the two curves of $y(x)$ (i.e.\ 
the modified Newtonian and Kerr cases) are practically indistinguishable. In 
order to amplify the tiny difference, we introduce $Y=f\cdot{y}$ where the 
normalization factor is given by $f=1-(x-1)/4$. The curves of $Y(x)$ are 
plotted in Fig.~\ref{fig1}b. We complement these graphs by showing 
(Fig.~\ref{fig1}c) $\delta=\sqrt{[\delta{x}]^2+[\delta{y}]^2}$ where 
$\delta{x(a)}$, $\delta{y(a)}$ are the differences in $x$ and $y$ between the 
modified Newtonian and Kerr cases. It can be seen that $\delta\lta0.25$, which 
indicates an accuracy of $x$ and $y$ better than about 20\%. The error is a 
maximum at $x\approx1.4$ and it goes sharply to zero for {\em{both\/}} $x=1$ 
($a=1$) and $x=3$ ($a=0$). One can construct analogous graphs for $\Phiabn$ but 
the results are very similar to those for $\Phi$. It is therefore a matter of 
taste which potential to choose for studying toroids in modified Newtonian 
potentials, but $\Phiabn$ is perhaps more practical as it has already been used 
by other authors (Miwa et al.\ 1998).

The structure of relativistic tori is determined by the radial acceleration 
along circular trajectories, and of course by the pressure gradient which, 
however, depends on the equation of state. We will therefore now discuss the 
radial acceleration for different $r={\it{const}}$ and different angular 
velocity $\omega$, and again we will compare the case of the modified Newtonian 
potential with that of the Kerr metric (free circular orbits in the equatorial 
plane have $\omega=1/(r^{3/2}+a)$ and acceleration magnitude $A=0$, but we do 
not restrict only to such cases). First, to explain how the graphs are 
constructed, we compare the acceleration for $\Phipw$ and for the Schwarzschild 
metric in Fig.~\ref{fig2}. Each curve gives the magnitude of the acceleration 
$A$ along $r=const$ orbits in the equatorial plane ($\theta=90\dg$) of the 
Schwarzschild metric. (Only the radial component contributes to the 
acceleration in the equatorial plane; a general expression valid also outside 
of the equatorial plane in Kerr spacetime was given explicitly by Semer\'ak 
1994). The radius progressively increases for the individual curves going from 
top to bottom of the plot. It has been widely discussed in the literature 
(Abramowicz \& Prasanna 1990) that $A(\omega)={\it{const}}$ at the photon 
orbit; this is indicated by a thick horizontal line in Fig.\ \ref{fig2}a. In 
fact, for applications to tori, we are mainly interested in orbits with radii 
greater than that of the marginally bound orbit, and therefore the whole 
portion of the graph corresponding to $r<\rmb$ is covered by shading. (In this 
region the modified Newtonian potential approach is not accurate.) One can 
compare the shape of the curves in the Schwarzschild case to the modified 
Newtonian case of $\Phipw$ in Fig.~\ref{fig2}b. The relative difference 
$\Delta$ between corresponding $A$'s from graphs \ref{fig2}a and \ref{fig2}b is 
plotted in the next two graphs, \ref{fig2}c--d, showing $\Delta(\omega)$ and 
$\Delta(w)$ ($w$ denotes the speed in the local frame of a non-rotating 
observer, which corresponds to angular velocity $\omega$; $-1<w<1$; $r>\rmb$). 
Here, the dimensionless quantity $\Delta$ is defined as

\beq
\Delta=\frac{A_{\rm{Exact}}(\omega)
 -A_{\rm{Modified\;Newtonian}}(\omega)}{A_{\rm{Exact}}(\omega=0)}
 \label{del}
\eeq
which is to be evaluated for fixed $r$, $\theta$ and $a$. The outermost
curve in Fig.\ \ref{fig2}c (with the largest magnitudes of $\Delta$)
corresponds to $r=\rmb$, while the innermost one (passing close to
$\omega=0$, $\Delta=0$) corresponds to $r=15\rh$.

Figure \ref{fig3} is constructed in the same way as Fig.\ \ref{fig2}, but
now it compares the Kerr $a=0.5$ case with the equivalent modified Newtonian
$\Phiabn$ case. By inspecting graphs with different $a$ we checked
that the accuracy of the modified Newtonian potentials $\Phipw$ and
$\Phiabn$ (as measured by $\Delta$) is comparable in the non-rotating
case but $\Phiabn$ is better as soon as $a$ is non-negligible. A similar
conclusion can be drawn for $\Phi$ from eq.\ (\ref{our}), and also for
circular orbits outside the equatorial plane. Analogous plots to those in
Figs.\ \ref{fig2}--\ref{fig3} have been constructed with other sets of
parameters. We find that acceptable accuracy of about 10\% in terms of
$\Delta$ is guaranteed whenever $r\gta1.5\rmb$.

\section{Conclusions} 
We have systematically checked and briefly illustrated that general 
relativistic effects on purely circular orbits can be imitated in a modified 
Newtonian potential. We have verified the accuracy of such models for the 
potentials $\Phi$ (eq.\ [\ref{our}]) and $\Phiabn$ (eq.\ [\ref{no5}]), finding 
that these two are comparable and that both offer higher accuracy than the 
usual Paczy\'nski-Wiita (1980) potential when the angular-momentum parameter 
$a$ is nonzero. By using our criterion concerning the relative accuracy of the 
acceleration along circular orbits, $|\Delta|\lta0.1$, we see that one can use 
the potential $\Phiabn$ of Artemova, Bj\"ornson \& Novikov (1996) 
satisfactorily for modelling tori in permanent orbital motion around a rotating 
black hole. The error increases very close to $\rmb$. The same conclusion holds 
for analogous potentials (such as the one proposed in this Note, 
eq.~[\ref{our}]) which reproduce the important orbits and accelerations for 
motion around a rotating black hole with an acceptable accuracy.

\ack
We greatly appreciate comments by an unknown referee; his/her advice helped us 
to improve presentation of our paper.
We acknowledge continued support from the project M\v{S}MT--Kontakt 
titled ``Spectral and Timing Properties of Cosmic Black Holes'' (LH14049), and 
the Czech Science Foundation grant ``Albert Einstein Center for Gravitation 
and Astrophysics'' (GA\v{C}R 14-37086G) in Prague. The Astronomical 
Institute of the Academy of Sciences has been operated under the project 
RVO:6798815.



\end{document}